\documentclass[conference]{IEEEtran}
\IEEEoverridecommandlockouts
\usepackage{cite}
\usepackage{amsmath,amssymb,amsfonts}
\usepackage{algorithmic}
\usepackage{graphicx}
\usepackage{caption}
\usepackage{showlabels}
\usepackage{textcomp}
\usepackage{xcolor}
\usepackage{hyperref}
\def\BibTeX{{\rm B\kern-.05em{\sc i\kern-.025em b}\kern-.08em
    T\kern-.1667em\lower.7ex\hbox{E}\kern-.125emX}}

\begin{document}

\title{Grasping Object: Challenges and Innovations in Robotics and Virtual Reality}

\author{\IEEEauthorblockN{Mingzhao Zhou*}
\IEEEauthorblockA{\textit{Department of Computer Science} \\
\textit{Brunel University London}\\
London, UK \\
mingzhao.zhou@brunel.ac.uk}
~\\
\and
\IEEEauthorblockN{Nadine Aburumman}
\IEEEauthorblockA{\textit{Department of Computer Science} \\
\textit{Brunel University London}\\
London, UK \\
nadine.aburumman@brunel.ac.uk}
}

\maketitle

\begin{abstract}
In real-life, grasping is one of the fundamental and effective forms of interaction when manipulating objects. This holds true in the physical and virtual world; however, unlike the physical world, virtual reality (VR) is grasped in a complex formulation that includes graphics, physics, and perception. In virtual reality, the user’s immersion level depends on realistic haptic feedback and high-quality graphics, which are computationally demanding and hard to achieve in real-time. Current solutions fail to produce plausible visuals and haptic feedback when simulation grasping in VR with a variety of targeted object dynamics. In this paper, we review the existing techniques for grasping in VR and robotics and indicate the main challenges that grasping faces in both domains. We aim to explore and understand the complexity of hand-grasping objects with different dynamics and inspire various ideas to improve and come up with potential solutions suitable for virtual reality applications.
\end{abstract}


\section{Introduction}

Over the past decades, with the advancement in virtual reality (VR) devices and Human-computer interaction (HCI) studies, the need for simulation of human behaviour with realistic interaction has become vital for many applications (see Fig. 1), such as industrial training, medical and surgical simulation, and rehabilitation \cite{shi2023, Qi2023, Manuela2019}. High-fidelity graphics in such virtual environments have been used to provide users with a relatively immersive virtual experience \cite{Mizuho2023}. However, obtaining a fully immersed experience requires realistic interaction with objects within the environment, where forces and masses of the objects can be felt during the interaction \cite{Jongyoon2022}. Haptic devices for both fingertips and/or the whole hand enable users to feel and manipulate the 3D objects and explore the virtual environment through a kinesthetic and cutaneous perception \cite{WANG2019}.

When the user interacts with an object, grasping is one of the main intuitive action behaviours. Although grasping behaviour is natural, human hands can grab a variety of objects with different shapes, weights, and frictions. Grasping in the virtual environment is a challenging task, as an ideal grasping action must take into account the geometry and dynamic characteristics of the virtual object \cite{Verschoor2018}. Despite the recent achievements in grasping techniques in VR, which have led to the emergence of some technological devices such as glove-based devices \cite{Liu2019} and controller-based devices \cite{CHoi2018, shi2023}. More technical methods need to be explored in order to ensure stable, controllable grasping in virtual reality.

The grasping techniques in VR differ broadly in complexity according to the virtual object's properties, such as mass, size, and materials. Moreover, the object's stiffness plays a significant role in making the interaction with the object complex to simulate \cite{Najdovski2018}. Most existing techniques focus on reducing the complexity of the grasping targets by simulating rigid bodies or objects with relatively simple and similar properties \cite{DELRIEU2020}. However, unlike rigid bodies, deformable bodies have high dynamic force attributes when the fingertips make contact with them. Therefore, obtaining stable grasping of deformable bodies while achieving realistic visuals and haptic feedback in virtual reality remains an open problem.
\begin{figure*}[ht]
\includegraphics[width=\textwidth]{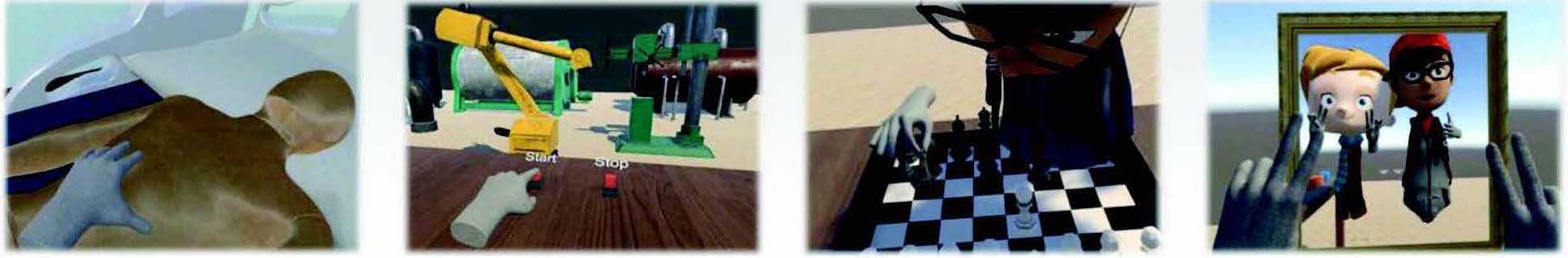}\captionof*{figure}{Fig 1.  Realistic interaction with virtual objects within an immersed environment is crucial for many applications, such
as industrial and medical training, entertainment and virtual social interaction \cite{Qi2023}.}
\end{figure*}
In this paper, we aim to review the existing methods for grasping from the perspectives of haptics and visuals, including the properties of the target objects. Further, to focus on the techniques of grasping in VR, we also generally introduce robotic grasping methods to compare the grasping methods in virtual environments with real-life robotics. Through this paper, we hope to summarise the main existing challenges in grasping simulation and improve the quality of VR grasping, potentially creating a direction for a new research agenda.

\section{VR techniques}

Objects with different shapes, masses, spatial positions and frictions can be grasped easily by human hands, which does not require too much effort \cite{Zhao2013}. Moreover, The process of realistic grasping is based on the laws of physics\cite{Borst2005}, according to \textit{Blaga et al.(2021)}, virtual objects are predominantly grasped with power, while real-world grasping is mainly based on precision\cite{Dalia2021}. However, to simulate the same behaviour in virtual reality, the virtual environment can not restrict the human hand motion according to physics laws like real environment\cite{Borst2005}, which may cause visual irrationality. To achieve realistic grasping in virtual reality, in recent decades, existing techniques have made great progress in visual feedback, which can render realistic objects \cite{Zhang2023},  while some types of technological devices have also been introduced to improve haptic feedback and rationalize interactions, such as haptic gloves, controllers, and haptic exoskeletons\cite{weichert2013, Hangxin2019, Ganias2023}. It is significant to understand the grasping mechanism, especially when it comes to tele-operations\cite{Milstein2021} and surgical operations\cite{WANG2017}. In this section, we will discuss virtual grasping methods which are centred around challenges and methods.

\subsection{Challenge}

Although grasping has made great progress in the development of current techniques, there are still challenges that remain in grasping simulation, especially in the simulation of precise grasping behaviour and grasping deformable bodies\cite{DELRIEU2020}. For large deformable bodies and heterogeneous deformation bodies\cite{Berenson2013}, grasping becomes more challenging. Some designs of grasping deformable bodies approximate objects to rigid bodies \cite{Oprea2019}, although this is limited. Most literature focuses on rigid bodies, as rigid bodies have relatively simple physical properties \cite{kyota2012}. Further, the current VR technique has a high demand for grasping arbitrary targets with the improvement of the virtual environment. Hence, it is significant to define grasping challenges and find effective solutions.

\subsubsection{High-Fidelity Modelling}

High-fidelity modelling is an essential and basic step in virtual grasping. One of the important models is the human hand. The whole hand has 27 DoF(degrees of freedom), and each finger except the thumb has three bones, which are distal, middle, and proximal phalanges, also these fingers have three joints, which are MCP (Metacarpophalangeal), PIP (Proximal Interphalangeal), and DIP (Distal Interphalangeal) joints(see Fig. 2, \cite{Burton2011}) \cite{Birouaș2020}.  Moreover, according to \textit{Mulatto et al.}, the human skill of manipulating objects depends on the thumb's kinematics and on its capability of opposing the other fingers. The thumb has 5 Dof, which allows humans to perform spatial movements\cite{Gabardi2018}. The complex features attract the attention of researchers. For a better simulation, \textit{Burton et al.(2011)} introduce a kinematic model of the human hand regarding the precise mapping of thumb movements\cite{Gabardi2018}. 
\textit{Kuch et al.(1995)} also present a hand model which can use the analysis-based approach to research the hand motion\cite{Kuch1995}.

\vspace{5mm} 
{
\centering
\includegraphics[width=0.45\textwidth]{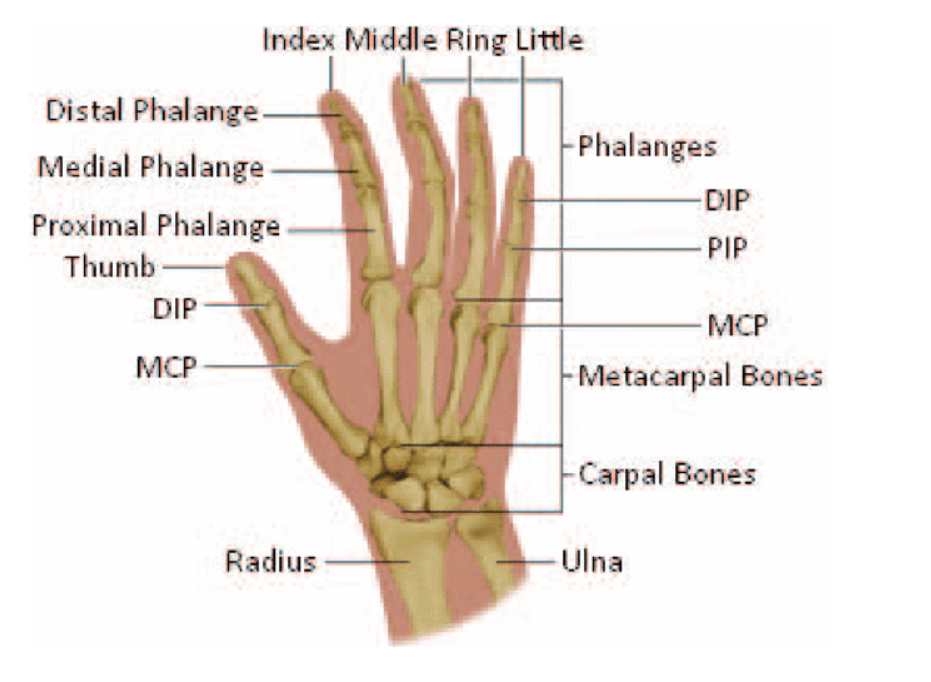}
\captionof*{figure}{Fig 2. The structure of the human hand skeleton, each finger except the thumb has three phalanges and joints\cite{Burton2011}.}
}
\vspace{5mm} 

Although we have been able to obtain relatively realistic models of human hands, it is hard to reproduce the surface stress based on the existing connection model between the human hand and the objects. How to model a precise and stable connection is still hard work\cite{DELRIEU2020}. Also, when it comes to free object manipulation, we need to consider the hand-bending range, one universal and effective way is collision detection. \textit{Oprea et al.(2019)} introduced a visually realistic grasping system, which allows users to interact with or manipulate non-predefined objects based on collision detection\cite{Oprea2019}. However, this method has penetration problems, which decreases the quality of the interaction.

Additionally, most researchers use virtual rigid fingers to simulate the interaction between the hand and the objects, according to \textit{Verschoor et al.(2018)}, using the soft fingers model would be an effective way to provide realistic feedback(see Fig 3, \cite{Verschoor2018}). However, using a high-resolution soft fingers model is limited by the high computational cost and realism. Other approaches use a part of the deformable skin, which is considered as a soft pad with a rigid skeleton \cite{Jacobs2011, Jacobs2012,Rumman2015}. The proposed method by \textit{Jacobs et al.(2011)} uses a new soft body model based on lattice shape matching (LSM)\cite{Jacobs2011}. Although they increase the robustness of the object manipulations, the soft pad may collapse in some cases. Moreover, The compliance between the bones and the tracker interpenetration results will be another factor we need to consider. In fact, existing soft finger modelling still faces many difficult problems to solve.

\vspace{5mm} 
{
\centering
\includegraphics[width=0.45\textwidth]{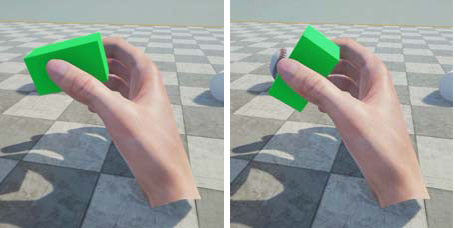}
\captionof*{figure}{Fig 3. Soft hand simulations while interacting with virtual objects within virtual reality, where the proposed method does not support self-collisions of the hand nor support contact between deformable objects \cite{Verschoor2018}.}
}
\vspace{5mm} 

\textbf{Visual Artifact and Overlapping:} As mentioned before, current VR techniques do not allow the virtual environment to constrain the real hand motion like the real environment. It will lead to unrealistic grasp behaviour and visually distracting artifacts\cite{Borst2005}, like the penetration between the hand models and virtual objects. The results have a bad performance in visual feedback, although we could keep the virtual hand model outside of the objects (see Fig 4). In addition, Hand-object overlapping is always a common problem in computer animation. It does not only happen in 3D objects grasping but also in 2D objects grasping\cite{Prachyabrued2012}. when the grasping targets have complex geometric features \cite{Boulic1996}, like vases, stuffed toys, and some handicrafts, the interaction process will lead to penetration \cite{Prachyabrued2016}. In some applications, due to cost reasons, the designers can not make detection in every frame, also, some video games more focus on bringing a fluent experience to users, which will ignore these computer animation problems. it is essential to find a low-cost method while avoiding penetration.

\vspace{5mm} 
{
\centering
\includegraphics[width=0.45\textwidth]{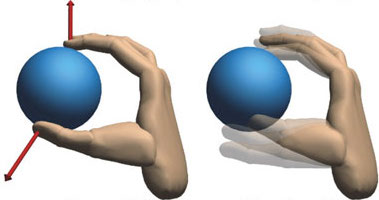}
\captionof*{figure}{Fig 4. Several grasping systems
included mechanisms to visually constrain a hand, which leads to artifact and delivery unrealistic experience \cite{Prachyabrued2016}.}
}
\vspace{5mm} 

\subsubsection{Device Performance}
 
VR devices in recent years have been improved in humanization and intelligence, which is different from early VR systems\cite {Doerner2022}.  The mainstream VR equipment includes visual HMD (Head-mounted display)and various types of haptic devices. Recently, The relatively famous visual display devices are HTC Vive series\cite{WikipediaHTC} and Oculus Quest series\cite{WikipediaOculusQuest}. Just in the past year, Vision Pro owned by Apple could provide users with a deep immersive experience\cite{WikipediaAppleVisionPro}. These devices have proved that MR(mixed-reality) can significantly improve users' immersive experience, which points the way for the development of virtual interaction. However, The considerations regarding the cost factor, process complexity, and safety factor become the obstacles which have blocked developers for many years. Especially the dexterity and portability of the device as well as battery life, which are problems that remain unsolved. (see Table 1).

\begin{table}[]
\centering
\begin{tabular}{|l|l|l|l|}
\hline
Devices      & Refresh rate    & Battery life & Mass     \\ \hline
Vision Pro   & 90Hz/96Hz/100Hz & 2.5 hours    & 600-650g \\ \hline
HTC Vive     & 90Hz            & 2 hours      & 555g     \\ \hline
Meta Quest 2 & 72Hz to 90Hz    & 2-3 hours    & 503g     \\ \hline
Meta Quest 3 & 90Hz            & 2-3 hours    & 512g     \\ \hline
\end{tabular}
\caption{Parameters of some popular devices, common problems faced are like short battery life and low portability\cite{WikipediaHTC, WikipediaQuest3, WikipediaOculusQuest, WikipediaAppleVisionPro}.}
\end{table}

Moreover, other problems have also been exposed, such as the screen door effect (see Figure 5\cite{Screen-qr}) and the barrel distortion (see Figure 6\cite{lens}). The screen door effect is a well-known problem that bothers users in the virtual environment. It means a visible grid-like pattern of thin lines or gaps between pixels on the screen, similar to a screen door. The reason is the resolution of current VR displays is not high enough to match the visual acuity of the human eye\cite{Anthes2016}. According to \textit{Nguyen et al.(2020)}, the solution of some commercial VR headsets is to use dispersive elements, which can effectively blur images on the screen\cite{Nguyen2020}. On the other hand, Barrel distortion is particularly likely to occur in zooming a planar image using a spherical mirror, the image magnification decreases with distance from the optical axis\cite{Anthes2016, lens}. To solve this problem, reversing transformation should be one solution, however, it will highly complex lens design\cite{Imagination}.

\vspace{5mm} 
{
\centering
\includegraphics[width=0.45\textwidth]{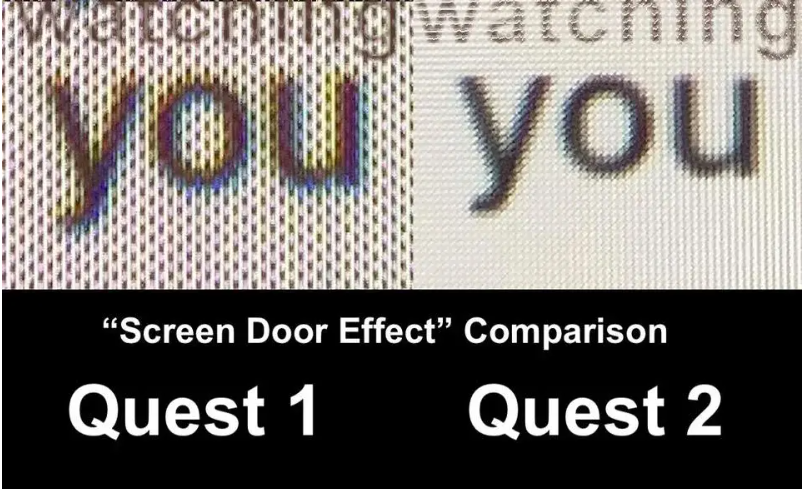}
\captionof*{figure}{Fig 5. The comparison regarding Screen Door Effect in Quest 1 and Quest 2\cite{Screen-qr}.}
}
\vspace{5mm} 

As we know, haptic devices have been widely developed by researchers, because the haptic can enhance the users' immersive experience\cite{Klein2021}. 
Equipment with relatively complete haptic feedback includes some haptic feedback gloves\cite{Wang2018}, haptic feedback exoskeleton\cite{Frisoli2005} and haptic devices like Touch Serious. These haptic devices tend to provide richer haptic feedback than controllers, which usually only provide force feedback. However, the devices either have difficulty delivering detailed renderings, are expensive to produce, or are not portable and wearable. Optimizing devices that can provide rich haptic feedback remains an important research direction.
In the following subsection, we will introduce VR grasping techniques from the conventional classification perspectives.

\vspace{5mm} 
{
\centering
\includegraphics[width=0.45\textwidth]{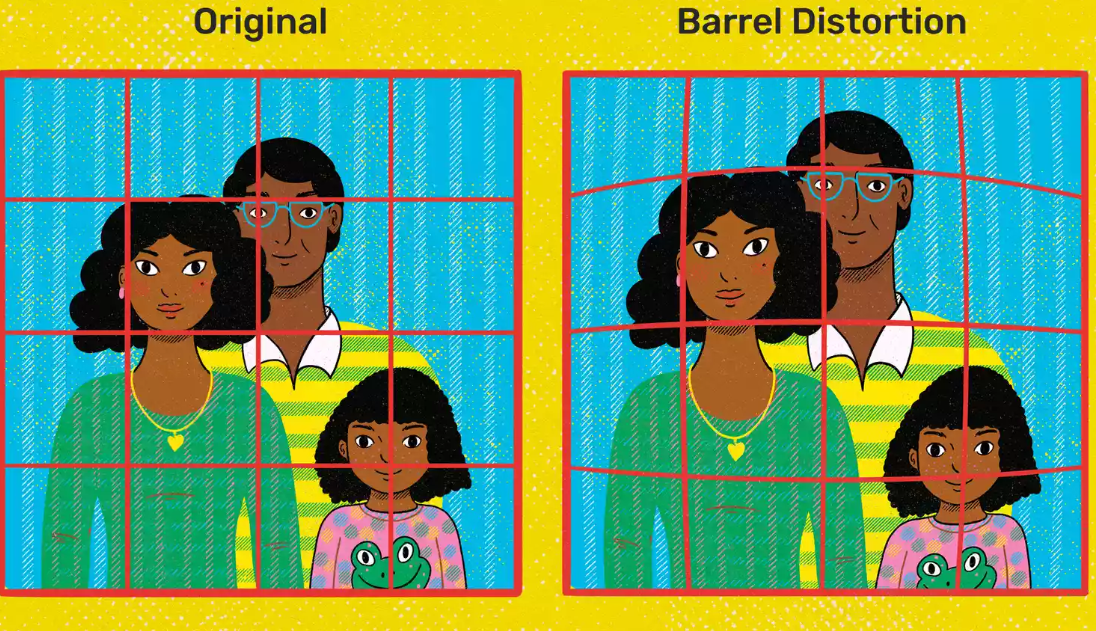}
\captionof*{figure}{Fig 6. The comparison between original graphic and Lens distortion graphic\cite{lens}.}
}
\vspace{5mm} 

{\subsubsection{Imperfect Theoretical Framework}
A complete theoretical system has not yet been formed in VR grasping. In particular, there is currently a lack of a common metric. In robotic grasping, \textit{Ferrari et al.(1992)} provide a well-known metric for grasp quality evaluation, which is Grasp Wrench Space(GWS). It refers to the set of all wrenches which can be applied to the object.\cite{Ferrari1992}. Through GWS, we can ensure the minimum distance between the origin and the boundary to explore the stability of the objects \cite{Zhang2022}. Moreover, within the robotic assembly research, The National Institute of Standards and Technology (NIST) presented a set of performance metrics and benchmarking tools, which contribute to technical specifications for robot systems \cite{Kimble2022-yd}. This benchmark can also address deformable bodies, however, to our knowledge, we can not find a common metric or benchmark in VR grasping, especially for grasping deformable bodies. Based on this, the motivation of this survey is to compare and discuss robotic grasping with VR grasping, and hope to provide a new research perspective for VR grasping.
}

\subsection{Grasping Methods}
\subsubsection{Physics-based Methods} 
Generally speaking, grasping in VR should focus on achieving stable and realistic interaction in geometry. Most existing methods use physical input devices to achieve grasping behaviour, which is constrained by the laws of physics, called the physics-based method\cite{GEORGII2005,nealen2006physically,moreno2022soft}. Specifically, when a 3D object model and a hand model are given, the VR system reacts to conscious action as a unique event by users touching off triggers or sensors, and then the target object can be grasped by a plan that generates hand object poses and hand joint angles to achieve stable grasp poses \cite{Remi2017}. This method has been well-researched in robotics; it solves the problems related to force attributes, like resistance to slipping and force closure. Coming to deformable bodies, this kind of method could be used to explain the interaction between the elastic surface and virtual fingertips (see Fig 7). \textit{Luo et al.(2007)} apply Bernoulli-Euler bending beam theory to the simulation of elastic shape deformation\cite{Luo2007}. This non-linear model could handle compliant motions with friction and complex changes in contact areas, but it is still limited to contiguous areas. To solve this problem, \textit{Tong et al.(2008)}introduce a non-linear contact force model and the beam-skeleton model for global shape deformation, which represent the multi-contact areas between the virtual hand and the deformable objects\cite{Tong2008}. However, it depends on the performance of the devices. Although existing motion-tracking devices could sense the hand in a precise way, when the objects come to small, complex shapes or high-deformation, the precision will be limited. 

\vspace{5mm} 
{
\centering
\includegraphics[width=0.5\textwidth]{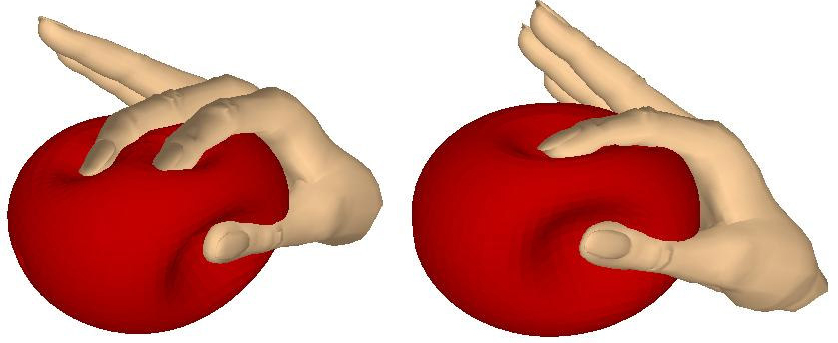}
\captionof*{figure}{Fig 7. Illustrates contact force and deformation
modelling for haptic simulation of grasping a deformable object
with a realistic virtual human hand \cite{Tong2008}, which runs at an interactive rate and needs efficient methods to be able to run in real-time.}
}
\vspace{5mm} 

To address this issue and improve the quality of the grasping, the performance of the devices is to be enhanced. Recently, some designs aimed to synchronise the virtual hand and tracking hand, as it was difficult for previous devices to ensure that the tracking devices coordinated and matched, which would lead to visual disharmony \cite{Buckingham2021}. The approach proposed by \textit{Delrieu et al.(2020)} enhances the existing technique by coupling the virtual kinematic hand with the visual hand tracking system in order to achieve precision grasping\cite{DELRIEU2020}, especially for small objects. In addition, providing hardware-based haptic feedback is also a good solution to optimise the users' experience \cite{Liu2019}. \textit{Jain et al.(2016)} introduce a VR system which could provide buoyancy, drag, and temperature changes\cite{Jain2016}. \textit{Liu et al.(2019)} use vibration motors to achieve the feedback on each finger\cite{Liu2019}.  \textit{Choi et al.(2016)} introduce a wearable haptic device to provide binary force feedback, which is brake-based\cite{Choi2016}. Although these devices do provide stable haptic feedback, as mentioned in the challenge, they always come with other drawbacks that cannot be ignored.

Force is the most significant physical constraint for haptic feedback, Simulating haptic feedback while grasping a 3D virtual object requires three-dimensional force/torque and a sufficient range of force magnitude to simulate contact forces during dexterous manipulation, and then render the power grasping of virtual objects \cite{mendes2019}. Moreover, to simulate subtle changes of contact forces between fingers and objects, we need to consider the role of high force resolution and dynamic response. In some tasks that require accurate force feedback, such as rotating a virtual tumour in surgical simulation, the error of the feedback force should be less than the human discrimination threshold of force size \cite {Bouzbib2023}. Thus, the user can infer the tumour's stiffness via the interaction between the resistance force and movement \cite{Misra2008}.
In the simulation of the force, Most researchers also take friction into account\cite{Klein2021}, but not gravity. However, reproducing gravity is significant for improving users' haptic feedback experience to be able to feel the weight of the virtual objects. Not only in the virtual world but also in the real world, gravity has important physical meanings, and simulating gravity significantly improves the virtual experience. Although haptic devices can not simulate gravity in VR, users are made aware of the objects that are grasped successfully through vibration (or forces) feedback \cite {Lopes2018}. A possible way to generate gravity is to change the mass of the device, like changing the liquid mass in dynamic circulation to achieve the simulation of gravity of different levels \cite {Rodolfo2022}. Such methods need bulky facilities and face high-cost problems. Skin stretch has recently been researched and could be used to express texture, friction, slip, and force. It can display the tangential force on the finger surface, which simulates gravity. Based on skin stretch, \textit{Choi et al.(2017)} designed a wearable haptic device that can simulate gravity to improve the grasping sensation\cite{Choi2017}.

\includegraphics[width=0.5\textwidth]{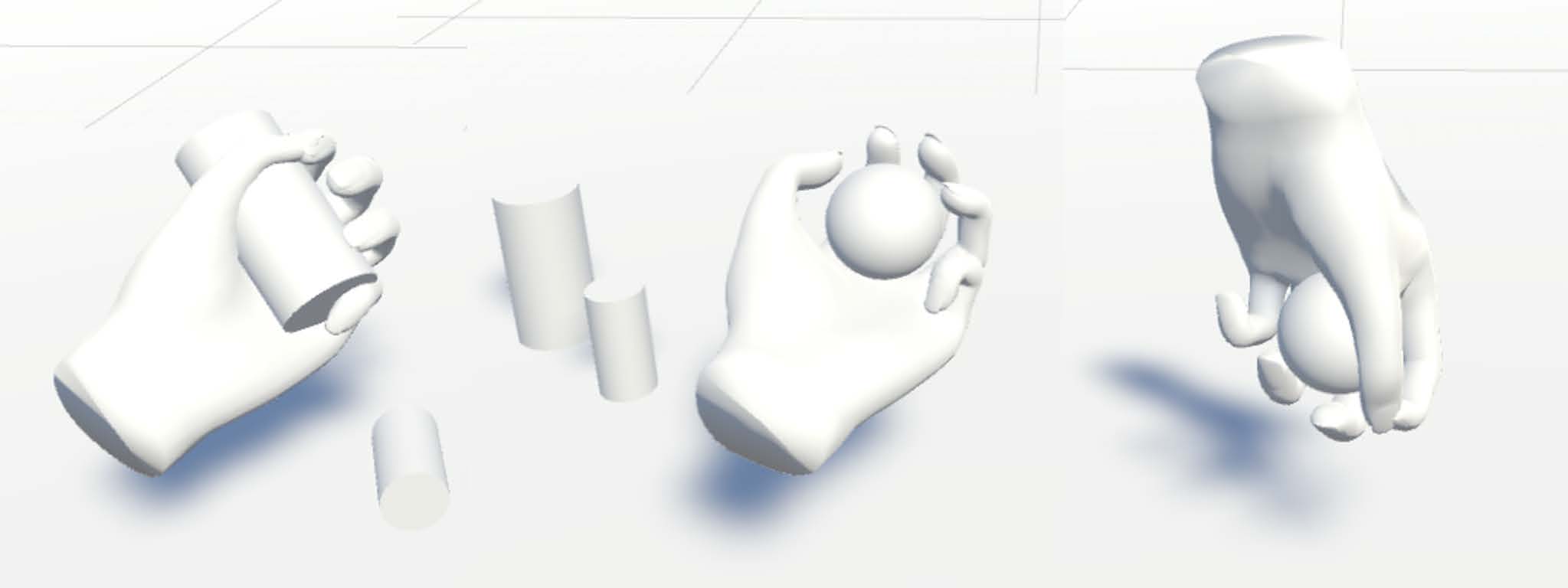}\captionof*{figure}{Fig 8. Shows grasping of handheld objects using geometric colliders, which employs GPU PhysX to detect collisions \cite{Friston2019}.}
\vspace{5mm} 

\textbf{Collision Detection}: Taking into account the collision between the object and the hand is another technique, which is called collision detection. Collision detection is the essential step of achieving realistic grasping, which refers to determining the intersection of two or more spatial targets. Although many existing gaming engines provide built-in collision detection algorithms, the complexity of the virtual environment in grasping sceneries requires stable and more efficient algorithms for VR. \textit{wang et al.(2023)} proposed an improved collision detection method, which is based on mesh simplification and particle swarm optimisation\cite{Wang2023}. this simplified method decreases the search space and improves the efficiency of collision detection. \textit{Macklin et al.(2014)} present a method using particles to treat contact and collisions \cite{Macklin2014} and \textit{Höll et al.(2018)} uses a simple ray-based technique according to a classic Coulomb contact model\cite{Höll2018} but only for an unconstrained hand-object interaction (see Fig. 8).
\vspace{5mm} 
\includegraphics[width=0.5\textwidth]{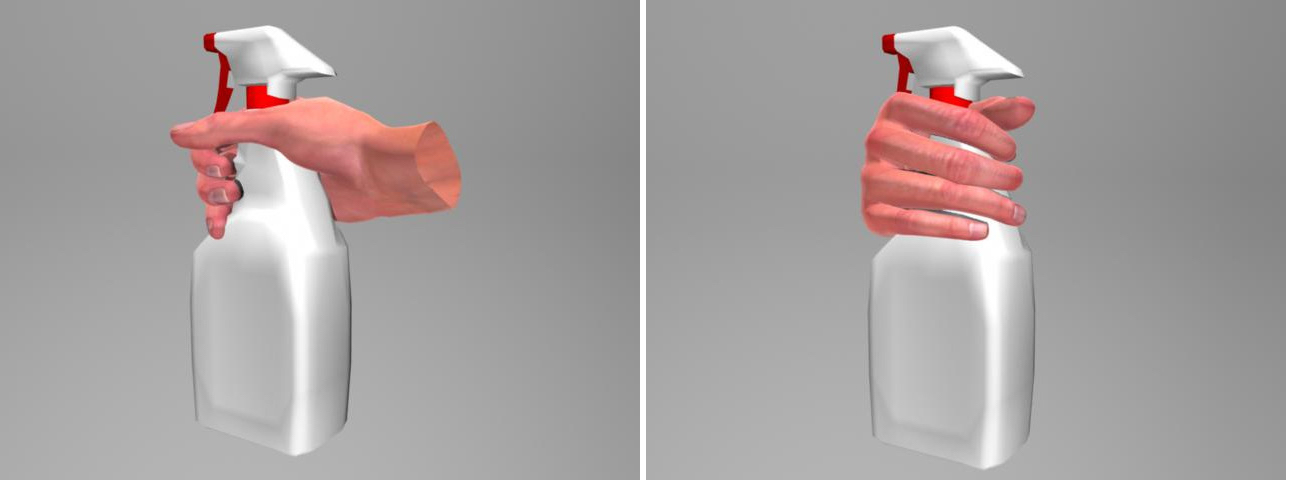}
\captionof*{figure}{Fig 9. Shows grasping of a complex virtual object, where the virtual hand matches the object geometry based on pre-defined poses \cite{Li2005}.}

\subsubsection{Other Visual Methods}
example-based animation could be able to produce compelling visual results during the interaction, which can significantly improve the quality of the experience. However, such approaches are limited to objects which have a predefined animation \cite{Oprea2019}. It means we should have prior knowledge regarding the grasping objects, especially for the deformable bodies, the transformation rules need to be established \cite{WANG2017}. In order to better achieve visual grasping, other strategies are like making the virtual object snap to the virtual hand \cite{Boulic1996}, the design proposed by \textit{Opera et al.(2019)} is a typical approach to snap the object to a virtual hand \cite{Oprea2019}. \textit{Nasim et al.(2016)} introduce an interesting method to solve complicated tasks, once the virtual hand collides with the objects, the hand will be deactivated. the concept is similar to 'freezing' \cite{Nasim2016}. Although these approaches can achieve relatively stable grasping, the interaction result will be unnatural, which limits dexterous operations and users' experiences.

Motivated by the manipulation of the 3D models in computer animation, the deformation transfer has also been used in grasping deformable bodies. Deformation transfer refers to using mesh sequences to represent 3D models. Generally, deformation gradients over triangle mesh transfer the deformation information from the source model to the target model \cite{muller2005meshless,CHEN2010,CASTI2019140}. This method applies to 3D models which have complex shapes like toys or handicrafts (see Fig.9). The relatively mature and popular deformation space-defined method is cage-based. However, the method presented by \cite{CHEN2010} requires building the cage manually for different targets, which is not user-friendly for novice users. \cite{Le2017} propose a cage-built method which could optimise the production of the initial cage. This interactive cage generation approach saves manual effort. Another method also introduces a nested cage-based method, which sets hierarchies at a step-by-step level, it is especially applicable to complex geometry and topology \cite{Sacht2015}. Cage-based approaches have been used to optimize natural interactions in VR. Especially for deformable bodies, cage-based deformation control could express the deformation in logic. Additionally, compared with another conventional method--force control, cage-based has a lower computational cost.

\subsubsection{Data-Driven Methods}
Physically based simulations can preserve plausibility by simulating interaction forces. However, such physical models must be driven by a controller \cite{Pollard2005}. In recent years, other methods like Data-based methods also make remarkable progress. This method records how a human performs grasping and then synthesises grasp motions to gain grasping knowledge. The advantage is that synthesized grasp poses are consistent with real word data, which are natural-looking. Moreover, the grasping data can be collected by digital gloves and optical devices, and the direct way to construct a grasp database is to collect grasping data from human volunteers. Obviously, this approach is time-consuming and cost-consuming for large-scale data acquisition. More and more researchers use data-driven methods to optimise grasping. \textit{Goldfeder et al.(2009)} introduce a grasp database which contains hundreds of thousands of form closure grasps for thousands of 3D models\cite{Goldfeder2009}.  \textit{Zacharias et al.(2009)} present object-specific grasp maps to encapsulate an object's manipulation characteristics\cite{Zacharias2009}. In addition, finding an appropriate grasp for an arbitrary object is a challenging problem, The grasping space needs to be considered with the DoF and geometry of the grasping targets, like \textit{Pelossof et al.(2004)} use machine learning to select an optimal grasp from the grasping space\cite{Pelossof2004}. 

Grasping rules were pre-defined for basic geometric primitives and could be used for other shapes. It works with databases which contain predefined poses, each one associated with an object geometry. Then, the most suitable grasp pose is selected according to the predefined grasp taxonomy or criteria.  \textit{Li et al.(2007)} introduce a shape-matching algorithm which matches hand shape to object shape by identifying collections of features, the matching is based on the similar relative placements and surface normals\cite{Li2007}. \textit{Pollard et al.(2005)} describe an approach which combines human motion data and physics-based simulation\cite{Pollard2005}. This method synthesizes the grasping and manipulating motions by incorporating physical constraints, which could decrease the visual artefacts, especially for shape differences. 
\textit{Tian et al.(2019)} present a real-time virtual grasping algorithm which computes the learned grasp space using machine learning and optimisation algorithms\cite{Tian2019}. The learned grasp space is precomputed using support vector machines and represents a set of stable grasp configurations that can be used to generate plausible grasp movements. 

However, such methods not only need plenty of time to build the dataset but also need to design excellent matching algorithms to save the time of traversing the dataset. Hence, improving the speed of indexing and matching is the direction in which the data-driven methods improve.

\section{Robotic Grasping}

Robotic grasping is a kind of realistic precision grasping, it is focused on solving specific tasks, which has a high demand for the matching ability and stability of the grasping system \cite{Kleeberger2020}. Further, Robotic grasping is vision-based grasping, which needs to keep stable contact in geometric. From the analytical heuristics strategies before decades to deep learning strategies in recent years, the complex scenes and tasks derive various grasping methods \cite{Souza2021}. In this section, we briefly introduce the robotic grasping methods to make discussion with VR grasping in the subsequent chapter. In addition, we recommend that readers refer to \cite{Zhang2022} for robotic grasping strategies in detail. 

\subsection{Analytic-based Methods}
The mainstream classification can be divided into two categories: analytic methods and data-driven methods. Basically, Analytic methods are applicable to situations in which the information of the target shape has been defined, with environment constraints to find the suitable grasping pose\cite{Kleeberger2020}. In the determining process of grasping, analytic methods need to be considered with the kinematics and dynamics formulations. Specifically speaking, based on the hand and object position, the grasping system defines the contact points of the object model and then synthesises the grasping, which is constrained by the selected criteria \cite{Sahbani2012}. 

Force-closure is a basic constraint \cite{Nguyen1986}, the contacts can apply an arbitrary wrench to the objects. When closure is generated, the objects can not move any more. the previous research was mainly focused on objects which have a sample shape or a finite number of flat faces, this case allows us to use the linear model to derive analytical formulation, for example, the position of a point on a face can be parameterized linearly by two variables, then it will be simplified as the linear problems \cite{Sahbani2012}.

It is worth mentioning that force closure does not guarantee stability. as the system is dynamic \cite{Nguyen1986}. According to \textit{Bicchi et al.(2000)} claim that the stability of the grasps is affected by the local properties of the geometry of the grasp and the force distribution, in addition to the locations of the contact points and the contact normals. Also, in non-force-closure grasps, the relative curvature becomes one of the determined factors in the contact grasp ability \cite{Bicchi2000}.

In recent years, caging-based techniques also have been used to solve the problem of grasping arbitrary objects. It refers to using the grippers to restrict the objects, which defines the grasping space through geometric calculations. \textit{Wan et al.(2012)} presents a method based on the caging principle to shrink fingers into immobilization \cite{Wan2012}. \textit{Kim et al.(2019)} also uses the caging-based method to grasp objects with deformable parts, They convert it to the 2D plane mathematical formula (see figure 10) in \cite{Kim2019}. \textit{Su et al.(2017)} focus on the caging grasping of the 3D objects, they construct the matrix which is represented by an attractive function to detect the boundary\cite{Su2017}. However, the gripper cage essentially is a rough configuration, which is not suitable for precise grasping tasks. One of the solutions is the heuristic algorithm, the algorithm will start from initial contacts, and then end once the selected grasp is force-closure\cite{Sahbani2012}. Heuristic algorithms also have been used to find optimality criteria, which refers to the algorithm that will search all the combinations, when the objects are modelled with a set of vertices \cite{Laguna2013}.

\subsection{Data-driven Methods}
With the development of machine learning, Data-driven methods would be another effective solution. Compared with the analytic-based method, the data-driven method avoids complex computation\cite{Bohg2014}. currently, there are two mainstream approaches to the data-driven method:

One is the system reproduces the grasps which are learned from the human operator. Researchers need to use devices to transfer data (like joint angles, hand shapes, wrench space etc.) to the workspace \cite{Sahbani2012}. For example, \textit{Aleotti et al.(2012)} propose a virtual reality system which enables the programming of robot grasps \cite{Aleotti2012}. The procedure allows the users to transfer the knowledge to the robotic system, which could be applied to the specific robotic device and objects. 

Another is the system learns the association between objects' characteristics and different hand shapes, in order to select the best grasps which fit the task objects \cite{Sahbani2012}. This approach can be generalized to new objects. \textit{Saxena et al.(2008)} present an algorithm which identifies characteristic points in each image to obtain a 3D grasping location. However, it can not fit some elongated objects, as the same grasping laws may not make realistic effects on all objects. Also, for the different objects, the algorithm may generate different grasp poses \cite{Saxena2008}. Moreover, \textit{Goldfeder et al.(2009)} construct a large grasping database, which uses the geometric similarity between objects, then generates force closure grasps (see figure11) \cite{Goldfeder2009}.

The data-driven methods, which according to previous experience in planning to grasp, also could be used to explain the interaction between the gripper and the deformable objects \cite{Souza2021}. \textit{Zaidi et al.(2017)} introduce a non-linear model, which presents the interaction between the deformable bodies and robotic hands\cite{Zaidi2017}. The complex simulation of the deformation has been possible. 

\vspace{5mm} 
{
\centering
\includegraphics[width=0.45\textwidth]{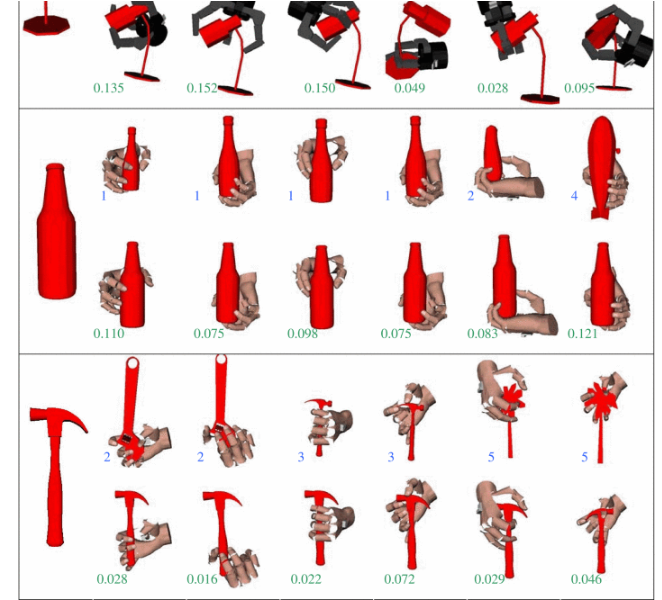}
\captionof*{figure}{Fig 11. The database developed by \textit{Goldfede et al.(2009)}, contains a large number of grasps \cite{Goldfeder2009}.}
}
\vspace{5mm} 

\vspace{5mm} 
{
\centering
\includegraphics[width=0.5\textwidth]{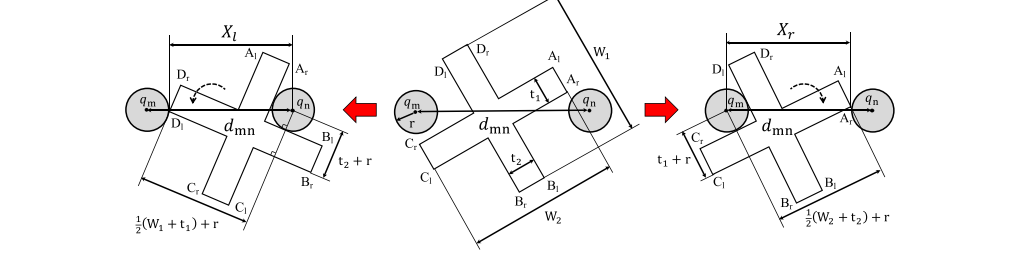}
\captionof*{figure}{Fig 10. \textit{Kim et al.(2019)} use mathematical formulas to derive the caging condition \cite{Kim2019}.}
}
\vspace{5mm} 
\textbf{Deformable and Fragile bodies}
As the existing techniques can achieve grasping in a stable way, the difficulty of tasks also becomes increasingly complex. Some specific tasks focus on the high-deformation even the fragile bodies, in these special cases, in addition to considering the interaction of the stability of objects with the robotic hand, it is also supposed to consider the deformation properties of the object itself. Most flexible deformable objects, basically, will go through three kinds of states: Elastic deformation, Plastic deformation, and Compressive failure \cite{Deformation-xi}. specifically, it is significant to avoid damage to the grasping targets. \textit{ Yu et al.(2023)} propose a method to predict deformation properties and the grasp pose of the objects \cite{Yu2023}, the method controls the deformation change in the range below the elastic limit, which could avoid the plastic deformation or damage of the grasping targets(orange, paper cup, etc.)(see figure 12 \cite{Yu2023}). \textit{Xu et al.(2020)} also try to solve the problem of damage and liquid spillage in the process of grasping hollow objects, like plastic bottles and paper cups\cite{Xu2020}. However, these methods are only applicable to some specific objects, and they are not versatile. 
For some special fragile objects, like tofu, grasping strategies could be more challenging. there is still no perfect way to solve this problem. One solution is to use fluid fingertips \cite{Adachi2015,Nishimura2016}. Another solution is to use micropatterned pads, the interesting method is motivated by the microstructure of the footpalm of some organisms in nature, like geckos. \textit{Nguyen et al.(2020)} refers to the foot palm of the tree frog to design the cushion of the gripper\cite{Nguyen2020}, which decreases the complex deformation analysis of fragile deformable objects during the grasping. 

\vspace{5mm} 
{
\centering
\includegraphics[width=0.45\textwidth]{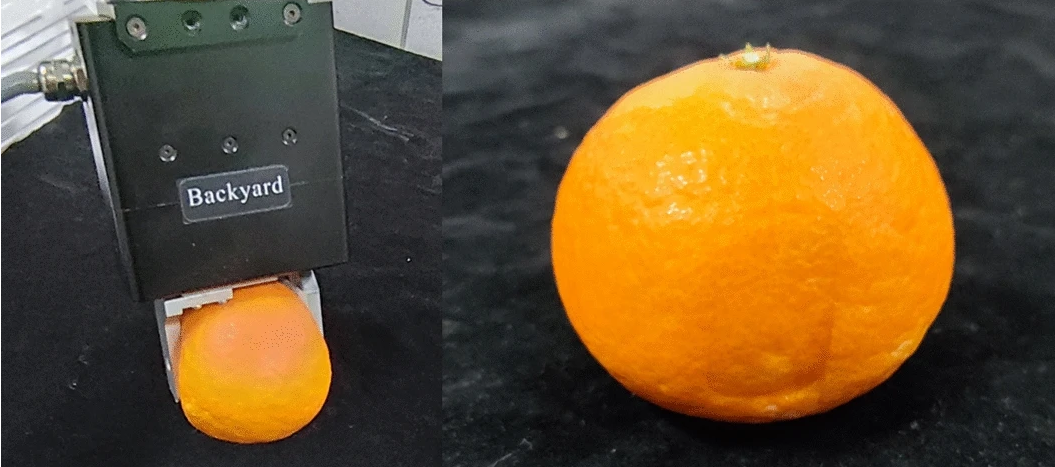}
\captionof*{figure}{Fig 12. Irreversible damage to the deformable object, such as orange, irreversible dents emerge on the surface of the peel\cite{Yu2023}.}
}
\vspace{5mm}

\section{Discussion}
\vspace{5mm} 
{
\centering
\includegraphics[width=0.5\textwidth]{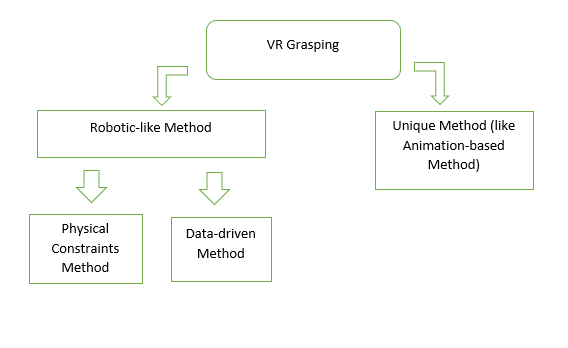}
\captionof*{figure}{Fig 13. A new virtual grasping classification based on robotic grasping.}
}
\vspace{5mm} 
From the above, we know humans hope that robotics could complete complex grasping tasks for them, especially for some grasping tasks involving safety issues and cost issues. The purpose of virtual grasping is to simulate the real world and try to bring the users a realistic virtual experience, which prefers to present situations that do not exist or are not easy to exist in life. Although the design directions are different, frankly speaking, virtual grasping solutions are similar to robotic grasping to a certain extent. Specifically, the design or improvement without prior knowledge in both robotics and virtual environments is inseparable from realistic physical laws. From this perspective, we can generate a grasping category: The physics-constraints method. The motivation for this grasping Strategy comes from human bare-hand grasping, which contributes to understanding the physical nature of the grasping. Nevertheless, human hand grasping is controlled by a complex system, the existing techniques can not achieve visual plausibility in handling unknown objects. Hence, Data-driven approaches are gaining traction. As machine learning has made great progress recently, the empirical method can be used in both robotics and virtual environments, in order to improve the stability and fluency of the interaction. Training on a large amount of data instead of planning with object models contributes to a new and effective development direction\cite{Zhang2022}. Data-driven methods substantially reduce the strong assumptions of the physical constraints methods, which makes grasping simulation give reasonable results when facing unknown objects. From Table 2, we can define the future grasping technology will be dominated by data-driven methods, combined with physical constraints to provide a versatility that conforms to the physical nature of human-hand grasping.

To better research VR grasping from a new perspective, based on robotic grasping, we classify the data-driven method and the physical constraints method into the category of robotic-like grasping. The advantage of this classification would be beneficial to VR researchers in understanding physical constraints from the relatively mature robotic grasping, such as force closure and force optimization. \textit{Nguyen(1986)} explained force closure grasp in detail\cite{Nguyen1986}, which is not only for robotics but also helps in VR. In addition, in terms of data-driven, virtual grasping can be optimized through machine grasping datasets to improve grasping visual fluency.

Unique virtual grasping strategies include animation-based methods, purely geometrical simulations and other visually feasible methods(see Figure 12). Taking into account the visual stability and randomness of the grasping targets, we think caging-based should be a potential solution for VR, which is also inspired by robotic caging grasping.  In robotic manipulation, caging grasping provides a way to manipulate an object without full immobilization and enables dealing with the pose uncertainties of the object. \textit{Kim et al.(2019)} and \textit{Su et al.(2019) } have demonstrated the cage-based approach is capable of grasping objects in a complex environment\cite{Kim2019, SU2019}. With the development of the grasping environment, various complex environments will be covered. Finding an approach that is versatile is essential. Cage-based points us in the direction, although it does not apply to precise grasping, it can achieve stable force closure, which is rational in visual. Additionally, in recent years, some data-driven-based methods also through approximate matching to achieve stable grasping as well.

\textit{Liu et al.(2019)} have already proposed that caging-based do have more positive performance in virtual grasping, compared with the conventional methods. They use collision geometry to find collision points, and then the virtual hand could hold the objects, although the grasp success rate will decrease when the targets come to complex bodies\cite{Liu2019}. The difficulty of approximation remains unresolved. For robotic grasping, the current problem is also how to approximate different objects with complex shapes to achieve caging. The caging-based grasping method still has a long way to go, however, it provides an effective strategy for arbitrary grasping targets in VR grasping, which avoids problems of other methods such as high cost, difficulty in handling with high deformation or requiring prior knowledge.

As mentioned before, caging-based deformation control has better performance than force-based control. The two caging-based are totally different concepts, the former refers to the grasping method, and the latter refers to the deformation transfer and control. Both methods are more competitive in their categories, especially for the deformable bodies. we think for VR further development, in addition to referring to the caging ideas in robot grasping, combing with the cage-based deformation control will be an effective solution for the grasping of deformable bodies.

\subsection{Computer Vision}
In virtual reality, we usually use visual feedback to ensure stable grasping. Even haptic feedback is designed to optimize the rationality of the interaction in visual scenery, as there is no lightweight device that can reproduce rich haptic sensations. In fact, the current VR research direction focuses on improving the naturalness and rationality of the grasping in visuals. Meanwhile, visual techniques are significant in robotic grasping, analysis methods detect collision points to achieve visually stable grasping. data-driven method estimates the pose and shape of 3D bodies from the depth image. Therefore, optimising fluency, rationality, and fidelity in visuals will be the key to users' immersive experience. From this perspective, virtual grasping cannot sidestep the techniques of computer vision (CV), like the tracking technique. Tracking is a continuous estimation of the position and orientation of an object\cite{Grimm20221}, which has achieved breakthroughs which can not be ignored. especially for hand tracking and hybrid tracking. Due to the high DOFs of the human hand and occlusion issues, hand tracking has been facing challenges. To address these problems, \textit{sharp et al.(2015)} provide a flexible hand tracking system, which avoids the limitations of close distance and glove-assisted devices. This system is based on machine learning to generate a large dataset for hand-pose hypotheses\cite{Sharp2015}. Another classic technique is feature-based tracking, this approach usually extracts edges or corners, which could be used for calculating the distance. For some visual features which have difficulty recognizing humans, corresponding feature detectors offer a quick and reliable solution\cite{Grimm20221}.

In recent decades, in addition to the classic methods mentioned above, some novel CV strategies are also beneficial to the development of VR, such as cloud-based tracking technique\cite{Openaai}, eye-tracking technique\cite{Clay2019-lr}. These technologies play an important role in enhancing human interaction with virtual objects. In fact, the development of VR is inseparable from the joint construction of multiple fields, Breakthroughs in such as the CV field can greatly promote the upgrade of VR interaction.

\section{Conclusion}
In this paper, towards the improvement of VR grasping and finding a novel solution, we reviewed the existing grasping techniques in the virtual environment and generally discussed the robotic grasping techniques.Finally, by comparison, it contributes to provide a new research perspective for VR grasping. Grasping has been studied for over 20 years, however, the simulation of grasping in the virtual environment still faces some great challenges from the perspective of the visuals and haptics. As the future virtual environment becomes increasingly complex, how to guarantee the users' immersive experience and avoid lags in interaction have become the upgrade problems of VR technology. Additionally, the solution regarding the challenges relies on the development of cross-disciplinary fields including materials, robotics, kinematics, computer vision and biology of the human behaviour.

\bibliographystyle{IEEEtran}
\bibliography{0.MainPaper}

\begin{thebibliography}{100}
\providecommand{\url}[1]{#1}
\csname url@samestyle\endcsname
\providecommand{\newblock}{\relax}
\providecommand{\bibinfo}[2]{#2}
\providecommand{\BIBentrySTDinterwordspacing}{\spaceskip=0pt\relax}
\providecommand{\BIBentryALTinterwordstretchfactor}{4}
\providecommand{\BIBentryALTinterwordspacing}{\spaceskip=\fontdimen2\font plus
\BIBentryALTinterwordstretchfactor\fontdimen3\font minus
  \fontdimen4\font\relax}
\providecommand{\BIBforeignlanguage}[2]{{%
\expandafter\ifx\csname l@#1\endcsname\relax
\typeout{** WARNING: IEEEtran.bst: No hyphenation pattern has been}%
\typeout{** loaded for the language `#1'. Using the pattern for}%
\typeout{** the default language instead.}%
\else
\language=\csname l@#1\endcsname
\fi
#2}}
\providecommand{\BIBdecl}{\relax}
\BIBdecl

\bibitem{shi2023}
\BIBentryALTinterwordspacing
Y.~Shi, L.~Zhao, X.~Lu, T.~Hoang, and M.~Wang, ``Grasping 3d objects with
  virtual hand in vr environment,'' in \emph{Proceedings of the 18th ACM
  SIGGRAPH International Conference on Virtual-Reality Continuum and Its
  Applications in Industry}, ser. VRCAI '22.\hskip 1em plus 0.5em minus
  0.4em\relax New York, NY, USA: Association for Computing Machinery, 2023.
  [Online]. Available: \url{https://doi.org/10.1145/3574131.3574428}
\BIBentrySTDinterwordspacing

\bibitem{Qi2023}
\BIBentryALTinterwordspacing
J.~Qi, F.~Gao, G.~Sun, J.~C. Yeo, and C.~T. Lim, ``Haptglove—untethered
  pneumatic glove for multimode haptic feedback in reality–virtuality
  continuum,'' \emph{Advanced Science}, vol.~10, no.~25, p. 2301044, 2023.
  [Online]. Available:
  \url{https://onlinelibrary.wiley.com/doi/abs/10.1002/advs.202301044}
\BIBentrySTDinterwordspacing

\bibitem{Manuela2019}
M.~Chessa, G.~Maiello, L.~K. Klein, V.~C. Paulun, and F.~Solari, ``Grasping
  objects in immersive virtual reality,'' in \emph{2019 IEEE Conference on
  Virtual Reality and 3D User Interfaces (VR)}, 2019, pp. 1749--1754.

\bibitem{Mizuho2023}
T.~Mizuho, T.~Narumi, and H.~Kuzuoka, ``Effects of the visual fidelity of
  virtual environments on presence, context-dependent forgetting, and
  source-monitoring error,'' \emph{IEEE Transactions on Visualization and
  Computer Graphics}, vol.~29, no.~5, pp. 2607--2614, 2023.

\bibitem{Jongyoon2022}
\BIBentryALTinterwordspacing
J.~Lim and Y.~Choi, ``Force-feedback haptic device for representation of tugs
  in virtual reality,'' \emph{Electronics}, vol.~11, no.~11, 2022. [Online].
  Available: \url{https://www.mdpi.com/2079-9292/11/11/1730}
\BIBentrySTDinterwordspacing

\bibitem{WANG2019}
\BIBentryALTinterwordspacing
D.~WANG, Y.~GUO, S.~LIU, Y.~ZHANG, W.~XU, and J.~XIAO, ``Haptic display for
  virtual reality: progress and challenges,'' \emph{Virtual Reality \&
  Intelligent Hardware}, vol.~1, no.~2, pp. 136--162, 2019, haptic Interaction.
  [Online]. Available:
  \url{https://www.sciencedirect.com/science/article/pii/S2096579619300130}
\BIBentrySTDinterwordspacing

\bibitem{Verschoor2018}
M.~Verschoor, D.~Lobo, and M.~A. Otaduy, ``Soft hand simulation for smooth and
  robust natural interaction,'' in \emph{2018 IEEE Conference on Virtual
  Reality and 3D User Interfaces (VR)}, 2018, pp. 183--190.

\bibitem{Liu2019}
H.~Liu, Z.~Zhang, X.~Xie, Y.~Zhu, Y.~Liu, Y.~Wang, and S.-C. Zhu,
  ``High-fidelity grasping in virtual reality using a glove-based system,'' in
  \emph{2019 International Conference on Robotics and Automation (ICRA)}, 2019,
  pp. 5180--5186.

\bibitem{CHoi2018}
\BIBentryALTinterwordspacing
I.~Choi, E.~Ofek, H.~Benko, M.~Sinclair, and C.~Holz, ``Claw: A multifunctional
  handheld haptic controller for grasping, touching, and triggering in virtual
  reality,'' in \emph{Proceedings of the 2018 CHI Conference on Human Factors
  in Computing Systems}, ser. CHI '18.\hskip 1em plus 0.5em minus 0.4em\relax
  New York, NY, USA: Association for Computing Machinery, 2018, p. 1–13.
  [Online]. Available: \url{https://doi.org/10.1145/3173574.3174228}
\BIBentrySTDinterwordspacing

\bibitem{Najdovski2018}
Z.~Najdovski and S.~Nahavandi, ``Extending haptic device capability for 3d
  virtual grasping,'' in \emph{Haptics: Perception, Devices and Scenarios},
  M.~Ferre, Ed.\hskip 1em plus 0.5em minus 0.4em\relax Berlin, Heidelberg:
  Springer Berlin Heidelberg, 2008, pp. 494--503.

\bibitem{DELRIEU2020}
T.~DELRIEU, V.~Weistroffer, and J.~P. Gazeau, ``Precise and realistic grasping
  and manipulation in virtual reality without force feedback,'' in \emph{2020
  IEEE Conference on Virtual Reality and 3D User Interfaces (VR)}, 2020, pp.
  266--274.

\bibitem{Zhao2013}
\BIBentryALTinterwordspacing
W.~Zhao, J.~Zhang, J.~Min, and J.~Chai, ``Robust realtime physics-based motion
  control for human grasping,'' \emph{ACM Trans. Graph.}, vol.~32, no.~6, nov
  2013. [Online]. Available: \url{https://doi.org/10.1145/2508363.2508412}
\BIBentrySTDinterwordspacing

\bibitem{Borst2005}
C.~Borst and A.~Indugula, ``Realistic virtual grasping,'' in \emph{IEEE
  Proceedings. VR 2005. Virtual Reality, 2005.}, 2005, pp. 91--98.

\bibitem{Dalia2021}
A.~Dalia~Blaga, M.~Frutos-Pascual, C.~Creed, and I.~Williams, ``A grasp on
  reality: Understanding grasping patterns for object interaction in real and
  virtual environments,'' in \emph{2021 IEEE International Symposium on Mixed
  and Augmented Reality Adjunct (ISMAR-Adjunct)}, 2021, pp. 391--396.

\bibitem{Zhang2023}
\BIBentryALTinterwordspacing
Z.~Zhang, L.~G. Giménez~Mateu, and J.~M. Fort, ``Apple vision pro: a new
  horizon in psychological research and therapy,'' \emph{Frontiers in
  Psychology}, vol.~14, 2023. [Online]. Available:
  \url{https://www.frontiersin.org/journals/psychology/articles/10.3389/fpsyg.2023.1280213}
\BIBentrySTDinterwordspacing

\bibitem{weichert2013}
F.~Weichert, D.~Bachmann, B.~Rudak, and D.~Fisseler, ``Analysis of the accuracy
  and robustness of the leap motion controller,'' \emph{Sensors}, vol.~13,
  no.~5, pp. 6380--6393, 2013.

\bibitem{Hangxin2019}
H.~Liu, Z.~Zhang, X.~Xie, Y.~Zhu, Y.~Liu, Y.~Wang, and S.-C. Zhu,
  ``High-fidelity grasping in virtual reality using a glove-based system,'' in
  \emph{2019 International Conference on Robotics and Automation (ICRA)}, 2019,
  pp. 5180--5186.

\bibitem{Ganias2023}
G.~Ganias, C.~Lougiakis, A.~Katifori, M.~Roussou, Y.~Ioannidis, and l.~P.
  Ioannidis, ``Comparing different grasping visualizations for object
  manipulation in vr using controllers,'' \emph{IEEE Transactions on
  Visualization and Computer Graphics}, vol.~29, no.~5, pp. 2369--2378, 2023.

\bibitem{Milstein2021}
A.~Milstein, L.~Alyagon, and I.~Nisky, ``Grip force control during virtual
  interaction with deformable and rigid objects via a haptic gripper,''
  \emph{IEEE Transactions on Haptics}, vol.~14, no.~3, pp. 564--576, 2021.

\bibitem{WANG2017}
\BIBentryALTinterwordspacing
Z.~Wang, M.~Fratarcangeli, A.~Ruimi, and A.~Srinivasa, ``Real time simulation
  of inextensible surgical thread using a kirchhoff rod model with force output
  for haptic feedback applications,'' \emph{International Journal of Solids and
  Structures}, vol. 113-114, pp. 192--208, 2017. [Online]. Available:
  \url{https://www.sciencedirect.com/science/article/pii/S0020768317300719}
\BIBentrySTDinterwordspacing

\bibitem{Berenson2013}
D.~Berenson, ``Manipulation of deformable objects without modeling and
  simulating deformation,'' in \emph{2013 IEEE/RSJ International Conference on
  Intelligent Robots and Systems}, 2013, pp. 4525--4532.

\bibitem{Oprea2019}
\BIBentryALTinterwordspacing
S.~Oprea, P.~Martinez-Gonzalez, A.~Garcia-Garcia, J.~A. Castro-Vargas,
  S.~Orts-Escolano, and J.~Garcia-Rodriguez, ``A visually realistic grasping
  system for object manipulation and interaction in virtual reality
  environments,'' \emph{Computers \& Graphics}, vol.~83, pp. 77--86, 2019.
  [Online]. Available:
  \url{https://www.sciencedirect.com/science/article/pii/S0097849319301098}
\BIBentrySTDinterwordspacing

\bibitem{kyota2012}
F.~Kyota and S.~Saito, ``Fast grasp synthesis for various shaped objects,'' in
  \emph{Computer graphics forum}, vol.~31, no. 2pt4.\hskip 1em plus 0.5em minus
  0.4em\relax Wiley Online Library, 2012, pp. 765--774.

\bibitem{Burton2011}
T.~M.~W. Burton, R.~Vaidyanathan, S.~C. Burgess, A.~J. Turton, and C.~Melhuish,
  ``Development of a parametric kinematic model of the human hand and a novel
  robotic exoskeleton,'' in \emph{2011 IEEE International Conference on
  Rehabilitation Robotics}, 2011, pp. 1--7.

\bibitem{Birouaș2020}
\BIBentryALTinterwordspacing
F.~I. Birouaș, R.~C. Țarcă, S.~Dzitac, and I.~Dzitac, ``Preliminary results
  in testing of a novel asymmetric underactuated robotic hand exoskeleton for
  motor impairment rehabilitation,'' \emph{Symmetry}, vol.~12, no.~9, 2020.
  [Online]. Available: \url{https://www.mdpi.com/2073-8994/12/9/1470}
\BIBentrySTDinterwordspacing

\bibitem{Gabardi2018}
M.~Gabardi, M.~Solazzi, D.~Leonardis, and A.~Frisoli, ``Design and evaluation
  of a novel 5 dof underactuated thumb-exoskeleton,'' \emph{IEEE Robotics and
  Automation Letters}, vol.~3, no.~3, pp. 2322--2329, 2018.

\bibitem{Kuch1995}
J.~Kuch and T.~Huang, ``Vision based hand modeling and tracking for virtual
  teleconferencing and telecollaboration,'' in \emph{Proceedings of IEEE
  International Conference on Computer Vision}, 1995, pp. 666--671.

\bibitem{Jacobs2011}
J.~Jacobs and B.~Froehlich, ``A soft hand model for physically-based
  manipulation of virtual objects,'' in \emph{2011 IEEE Virtual Reality
  Conference}, 2011, pp. 11--18.

\bibitem{Jacobs2012}
J.~Jacobs, M.~Stengel, and B.~Froehlich, ``A generalized god-object method for
  plausible finger-based interactions in virtual environments,'' in \emph{2012
  IEEE Symposium on 3D User Interfaces (3DUI)}, 2012, pp. 43--51.

\bibitem{Rumman2015}
\BIBentryALTinterwordspacing
N.~A. Rumman, M.~Schaerf, and D.~Bechmann, ``Collision detection for
  articulated deformable characters,'' in \emph{Proceedings of the 8th ACM
  SIGGRAPH Conference on Motion in Games}, ser. MIG '15.\hskip 1em plus 0.5em
  minus 0.4em\relax New York, NY, USA: Association for Computing Machinery,
  2015, p. 215–220. [Online]. Available:
  \url{https://doi.org/10.1145/2822013.2822034}
\BIBentrySTDinterwordspacing

\bibitem{Prachyabrued2012}
\BIBentryALTinterwordspacing
M.~Prachyabrued and C.~W. Borst, ``Virtual grasp release method and
  evaluation,'' \emph{Int. J. Hum.-Comput. Stud.}, vol.~70, no.~11, p.
  828–848, nov 2012. [Online]. Available:
  \url{https://doi.org/10.1016/j.ijhcs.2012.06.002}
\BIBentrySTDinterwordspacing

\bibitem{Boulic1996}
\BIBentryALTinterwordspacing
R.~Boulic, S.~Rezzonico, and D.~Thalmann, ``Multi-finger manipulation of
  virtual objects,'' in \emph{Proceedings of the ACM Symposium on Virtual
  Reality Software and Technology}, ser. VRST '96.\hskip 1em plus 0.5em minus
  0.4em\relax New York, NY, USA: Association for Computing Machinery, 1996, p.
  67–74. [Online]. Available: \url{https://doi.org/10.1145/3304181.3304195}
\BIBentrySTDinterwordspacing

\bibitem{Prachyabrued2016}
M.~Prachyabrued and C.~W. Borst, ``Design and evaluation of visual
  interpenetration cues in virtual grasping,'' \emph{IEEE Transactions on
  Visualization and Computer Graphics}, vol.~22, no.~6, pp. 1718--1731, 2016.

\bibitem{Doerner2022}
\BIBentryALTinterwordspacing
R.~Doerner, W.~Broll, B.~Jung, P.~Grimm, M.~G{\"o}bel, and R.~Kruse,
  \emph{Introduction to Virtual and Augmented Reality}.\hskip 1em plus 0.5em
  minus 0.4em\relax Cham: Springer International Publishing, 2022, pp. 1--37.
  [Online]. Available: \url{https://doi.org/10.1007/978-3-030-79062-2_1}
\BIBentrySTDinterwordspacing

\bibitem{WikipediaHTC}
{Wikipedia contributors}, ``{HTC} vive,''
  \url{https://en.wikipedia.org/w/index.php?title=HTC_Vive & oldid=1193839847},
  Jan. 2024, accessed: NA-NA-NA.

\bibitem{WikipediaOculusQuest}
{Wikipedia contributor}, ``Oculus quest,''
  \url{https://en.wikipedia.org/w/index.php?title=Oculus_Quest \&
  oldid=1220416255}, Apr. 2024, accessed: NA-NA-NA.

\bibitem{WikipediaAppleVisionPro}
{Wikipedia contributors}, ``Apple vision pro,''
  \url{https://en.wikipedia.org/w/index.php?title=Apple_Vision_Pro\&
  oldid=1221462128}, Apr. 2024, accessed: NA-NA-NA.

\bibitem{WikipediaQuest3}
{Wikipedia contributor}, ``Meta quest 3,''
  \url{https://en.wikipedia.org/w/index.php?title=Meta_Quest_3 \&
  oldid=1220416361}, Apr. 2024, accessed: NA-NA-NA.

\bibitem{Screen-qr}
M.~Richardson, ``\BIBforeignlanguage{en}{What is: Screen door effect, mura,
  aliasing in {VR}},''
  \url{https://www.hlplanet.com/screen-door-effect-mura-aliasing/}, Dec. 2023,
  accessed: 2024-4-30.

\bibitem{lens}
J.~Plumridge, ``How to fix lens barrel distortion in the camera,''
  \url{https://www.lifewire.com/what-is-barrel-lens-distortion-493725}, Jul.
  2011, accessed: 2024-4-30.

\bibitem{Anthes2016}
C.~Anthes, R.~J. García-Hernández, M.~Wiedemann, and D.~Kranzlmüller,
  ``State of the art of virtual reality technology,'' in \emph{2016 IEEE
  Aerospace Conference}, 2016, pp. 1--19.

\bibitem{Nguyen2020}
P.~Van~Nguyen, Q.~K. Luu, Y.~Takamura, and V.~A. Ho, ``Wet adhesion of
  micro-patterned interfaces for stable grasping of deformable objects,'' in
  \emph{2020 IEEE/RSJ International Conference on Intelligent Robots and
  Systems (IROS)}, 2020, pp. 9213--9219.

\bibitem{Imagination}
{Imagination Technologies}, ``\BIBforeignlanguage{en}{Speeding up {GPU} barrel
  distortion correction in mobile {VR} - imagination},''
  \url{https://blog.imaginationtech.com/speeding-up-gpu-barrel-distortion-correction-in-mobile-vr/},
  accessed: 2024-4-30.

\bibitem{Klein2021}
L.~Klein, G.~Maiello, R.~Fleming, and D.~Voudouris, ``Friction is preferred
  over grasp configuration in precision grip grasping,'' \emph{Journal of
  Neurophysiology}, vol. 125, 02 2021.

\bibitem{Wang2018}
D.~Wang, M.~Song, N.~Afzal, Y.~Zheng, W.~Xu, and Y.~Zhang, ``Toward whole-hand
  kinesthetic feedback: A survey of force feedback gloves,'' \emph{IEEE
  Transactions on Haptics}, vol.~PP, pp. 1--1, 11 2018.

\bibitem{Frisoli2005}
A.~Frisoli, F.~Rocchi, S.~Marcheschi, A.~Dettori, F.~Salsedo, and
  M.~Bergamasco, ``A new force-feedback arm exoskeleton for haptic interaction
  in virtual environments,'' in \emph{First Joint Eurohaptics Conference and
  Symposium on Haptic Interfaces for Virtual Environment and Teleoperator
  Systems. World Haptics Conference}, 2005, pp. 195--201.

\bibitem{Ferrari1992}
C.~Ferrari and J.~Canny, ``Planning optimal grasps,'' in \emph{Proceedings 1992
  IEEE International Conference on Robotics and Automation}, 1992, pp.
  2290--2295 vol.3.

\bibitem{Zhang2022}
H.~Zhang, J.~Tang, S.~Sun, and X.~Lan, ``Robotic grasping from classical to
  modern: A survey,'' 02 2022.

\bibitem{Kimble2022-yd}
K.~Kimble, J.~Albrecht, M.~Zimmerman, and J.~Falco,
  ``\BIBforeignlanguage{en}{Performance measures to benchmark the grasping,
  manipulation, and assembly of deformable objects typical to manufacturing
  applications},'' \emph{\BIBforeignlanguage{en}{Front. Robot. AI}}, vol.~9, p.
  999348, Nov. 2022.

\bibitem{GEORGII2005}
\BIBentryALTinterwordspacing
J.~Georgii and R.~Westermann, ``Mass-spring systems on the gpu,''
  \emph{Simulation Modelling Practice and Theory}, vol.~13, no.~8, pp.
  693--702, 2005, programmable Graphics Hardware. [Online]. Available:
  \url{https://www.sciencedirect.com/science/article/pii/S1569190X05000857}
\BIBentrySTDinterwordspacing

\bibitem{nealen2006physically}
A.~Nealen, M.~M{\"u}ller, R.~Keiser, E.~Boxerman, and M.~Carlson, ``Physically
  based deformable models in computer graphics,'' in \emph{Computer graphics
  forum}, vol.~25, no.~4.\hskip 1em plus 0.5em minus 0.4em\relax Wiley Online
  Library, 2006, pp. 809--836.

\bibitem{moreno2022soft}
M.~R. Moreno-Guerra, O.~Mart{\'\i}nez-Romero, L.~M. Palacios-Pineda,
  D.~Olvera-Trejo, J.~A. Diaz-Elizondo, E.~Flores-Villalba, J.~V. da~Silva,
  A.~El{\'\i}as-Z{\'u}{\~n}iga, and C.~A. Rodriguez, ``Soft tissue hybrid model
  for real-time simulations,'' \emph{Polymers}, vol.~14, no.~7, p. 1407, 2022.

\bibitem{Remi2017}
F.~J.~V. Remi~Alkemade and S.~G. Lukosch, ``On the efficiency of a vr hand
  gesture-based interface for 3d object manipulations in conceptual design,''
  \emph{International Journal of Human–Computer Interaction}, vol.~33,
  no.~11, pp. 882--901, 2017.

\bibitem{Luo2007}
Q.~Luo and J.~Xiao, ``Contact and deformation modeling for interactive
  environments,'' \emph{IEEE Transactions on Robotics}, vol.~23, no.~3, pp.
  416--430, 2007.

\bibitem{Tong2008}
T.~Cui, J.~Xiao, and A.~Song, ``Simulation of grasping deformable objects with
  a virtual human hand,'' in \emph{2008 IEEE/RSJ International Conference on
  Intelligent Robots and Systems}, 2008, pp. 3965--3970.

\bibitem{Buckingham2021}
\BIBentryALTinterwordspacing
G.~Buckingham, ``Hand tracking for immersive virtual reality: Opportunities and
  challenges,'' \emph{Frontiers in Virtual Reality}, vol.~2, 2021. [Online].
  Available:
  \url{https://www.frontiersin.org/articles/10.3389/frvir.2021.728461}
\BIBentrySTDinterwordspacing

\bibitem{Jain2016}
\BIBentryALTinterwordspacing
D.~Jain, M.~Sra, J.~Guo, R.~Marques, R.~Wu, J.~Chiu, and C.~Schmandt,
  ``Immersive terrestrial scuba diving using virtual reality,'' in
  \emph{Proceedings of the 2016 CHI Conference Extended Abstracts on Human
  Factors in Computing Systems}, ser. CHI EA '16.\hskip 1em plus 0.5em minus
  0.4em\relax New York, NY, USA: Association for Computing Machinery, 2016, p.
  1563–1569. [Online]. Available:
  \url{https://doi.org/10.1145/2851581.2892503}
\BIBentrySTDinterwordspacing

\bibitem{Choi2016}
I.~Choi, E.~W. Hawkes, D.~L. Christensen, C.~J. Ploch, and S.~Follmer,
  ``Wolverine: A wearable haptic interface for grasping in virtual reality,''
  in \emph{2016 IEEE/RSJ International Conference on Intelligent Robots and
  Systems (IROS)}, 2016, pp. 986--993.

\bibitem{mendes2019}
D.~Mendes, F.~M. Caputo, A.~Giachetti, A.~Ferreira, and J.~Jorge, ``A survey on
  3d virtual object manipulation: From the desktop to immersive virtual
  environments,'' in \emph{Computer graphics forum}, vol.~38, no.~1.\hskip 1em
  plus 0.5em minus 0.4em\relax Wiley Online Library, 2019, pp. 21--45.

\bibitem{Bouzbib2023}
E.~Bouzbib, C.~Pacchierotti, and A.~Lécuyer, ``When tangibles become
  deformable: Studying pseudo-stiffness perceptual thresholds in a vr grasping
  task,'' \emph{IEEE Transactions on Visualization and Computer Graphics},
  vol.~29, no.~5, pp. 2743--2752, 2023.

\bibitem{Misra2008}
\BIBentryALTinterwordspacing
S.~Misra, K.~T. Ramesh, and A.~M. Okamura, ``Modeling of tool-tissue
  interactions for computer-based surgical simulation: A literature review,''
  \emph{Presence: Teleoper. Virtual Environ.}, vol.~17, no.~5, p. 463–491,
  oct 2008. [Online]. Available: \url{https://doi.org/10.1162/pres.17.5.463}
\BIBentrySTDinterwordspacing

\bibitem{Lopes2018}
\BIBentryALTinterwordspacing
P.~Lopes, S.~You, A.~Ion, and P.~Baudisch, ``Adding force feedback to mixed
  reality experiences and games using electrical muscle stimulation,'' in
  \emph{Proceedings of the 2018 CHI Conference on Human Factors in Computing
  Systems}, ser. CHI '18.\hskip 1em plus 0.5em minus 0.4em\relax New York, NY,
  USA: Association for Computing Machinery, 2018, p. 1–13. [Online].
  Available: \url{https://doi.org/10.1145/3173574.3174020}
\BIBentrySTDinterwordspacing

\bibitem{Rodolfo2022}
\BIBentryALTinterwordspacing
R.~Garc\'{\i}a-Rodr\'{\i}guez, V.~Segovia-Palacios, V.~Parra-Vega, and
  M.~Villalva-Lucio, ``Dynamic optimal grasping of a circular object with
  gravity using robotic soft-fingertips,'' \emph{Int. J. Appl. Math. Comput.
  Sci.}, vol.~26, no.~2, p. 309–323, jun 2016. [Online]. Available:
  \url{https://doi.org/10.1515/amcs-2016-0022}
\BIBentrySTDinterwordspacing

\bibitem{Choi2017}
\BIBentryALTinterwordspacing
I.~Choi, H.~Culbertson, M.~R. Miller, A.~Olwal, and S.~Follmer, ``Grabity: A
  wearable haptic interface for simulating weight and grasping in virtual
  reality,'' in \emph{Proceedings of the 30th Annual ACM Symposium on User
  Interface Software and Technology}, ser. UIST '17.\hskip 1em plus 0.5em minus
  0.4em\relax New York, NY, USA: Association for Computing Machinery, 2017, p.
  119–130. [Online]. Available: \url{https://doi.org/10.1145/3126594.3126599}
\BIBentrySTDinterwordspacing

\bibitem{Friston2019}
S.~Friston, E.~Griffith, D.~Swapp, A.~Marshall, and A.~Steed, ``Position-based
  control of under-constrained haptics: A system for the dexmo glove,''
  \emph{IEEE Robotics and Automation Letters}, vol.~4, no.~4, pp. 3497--3504,
  2019.

\bibitem{Wang2023}
T.~Wang, ``Collision detection algorithm based on particle system in virtual
  simulation,'' in \emph{2023 4th International Conference for Emerging
  Technology (INCET)}, 2023, pp. 1--5.

\bibitem{Macklin2014}
\BIBentryALTinterwordspacing
M.~Macklin, M.~M\"{u}ller, N.~Chentanez, and T.-Y. Kim, ``Unified particle
  physics for real-time applications,'' \emph{ACM Trans. Graph.}, vol.~33,
  no.~4, jul 2014. [Online]. Available:
  \url{https://doi.org/10.1145/2601097.2601152}
\BIBentrySTDinterwordspacing

\bibitem{Höll2018}
M.~Höll, M.~Oberweger, C.~Arth, and V.~Lepetit, ``Efficient physics-based
  implementation for realistic hand-object interaction in virtual reality,'' in
  \emph{2018 IEEE Conference on Virtual Reality and 3D User Interfaces (VR)},
  2018, pp. 175--182.

\bibitem{Li2005}
Y.~Li and N.~Pollard, ``A shape matching algorithm for synthesizing humanlike
  enveloping grasps,'' in \emph{5th IEEE-RAS International Conference on
  Humanoid Robots, 2005.}, 2005, pp. 442--449.

\bibitem{Nasim2016}
\BIBentryALTinterwordspacing
K.~Nasim and Y.~J. Kim, ``Physics-based interactive virtual grasping,'' in
  \emph{Proceedings of HCI Korea}, ser. HCIK '16.\hskip 1em plus 0.5em minus
  0.4em\relax Seoul, KOR: Hanbit Media, Inc., 2016, p. 114–120. [Online].
  Available: \url{https://doi.org/10.17210/hcik.2016.01.114}
\BIBentrySTDinterwordspacing

\bibitem{muller2005meshless}
M.~M{\"u}ller, B.~Heidelberger, M.~Teschner, and M.~Gross, ``Meshless
  deformations based on shape matching,'' \emph{ACM transactions on graphics
  (TOG)}, vol.~24, no.~3, pp. 471--478, 2005.

\bibitem{CHEN2010}
\BIBentryALTinterwordspacing
L.~Chen, J.~Huang, H.~Sun, and H.~Bao, ``Cage-based deformation transfer,''
  \emph{Computers \& Graphics}, vol.~34, no.~2, pp. 107--118, 2010. [Online].
  Available:
  \url{https://www.sciencedirect.com/science/article/pii/S0097849310000208}
\BIBentrySTDinterwordspacing

\bibitem{CASTI2019140}
\BIBentryALTinterwordspacing
S.~Casti, M.~Livesu, N.~Mellado, N.~{Abu Rumman}, R.~Scateni, L.~Barthe, and
  E.~Puppo, ``Skeleton based cage generation guided by harmonic fields,''
  \emph{Computers \& Graphics}, vol.~81, pp. 140--151, 2019. [Online].
  Available:
  \url{https://www.sciencedirect.com/science/article/pii/S0097849319300457}
\BIBentrySTDinterwordspacing

\bibitem{Le2017}
\BIBentryALTinterwordspacing
B.~H. Le and Z.~Deng, ``Interactive cage generation for mesh deformation,'' in
  \emph{Proceedings of the 21st ACM SIGGRAPH Symposium on Interactive 3D
  Graphics and Games}, ser. I3D '17.\hskip 1em plus 0.5em minus 0.4em\relax New
  York, NY, USA: Association for Computing Machinery, 2017. [Online].
  Available: \url{https://doi.org/10.1145/3023368.3023369}
\BIBentrySTDinterwordspacing

\bibitem{Sacht2015}
\BIBentryALTinterwordspacing
L.~Sacht, E.~Vouga, and A.~Jacobson, ``Nested cages,'' \emph{ACM Trans.
  Graph.}, vol.~34, no.~6, nov 2015. [Online]. Available:
  \url{https://doi.org/10.1145/2816795.2818093}
\BIBentrySTDinterwordspacing

\bibitem{Pollard2005}
\BIBentryALTinterwordspacing
N.~S. Pollard and V.~B. Zordan, ``Physically based grasping control from
  example,'' in \emph{Proceedings of the 2005 ACM SIGGRAPH/Eurographics
  Symposium on Computer Animation}, ser. SCA '05.\hskip 1em plus 0.5em minus
  0.4em\relax New York, NY, USA: Association for Computing Machinery, 2005, p.
  311–318. [Online]. Available: \url{https://doi.org/10.1145/1073368.1073413}
\BIBentrySTDinterwordspacing

\bibitem{Goldfeder2009}
C.~Goldfeder, M.~Ciocarlie, H.~Dang, and P.~K. Allen, ``The columbia grasp
  database,'' in \emph{2009 IEEE International Conference on Robotics and
  Automation}, 2009, pp. 1710--1716.

\bibitem{Zacharias2009}
F.~Zacharias, C.~Borst, and G.~Hirzinger, ``Object-specific grasp maps for use
  in planning manipulation actions,'' \emph{German Workshop on Robotics}, pp.
  203--213, 01 2009.

\bibitem{Pelossof2004}
R.~Pelossof, A.~Miller, P.~Allen, and T.~Jebara, ``An svm learning approach to
  robotic grasping,'' in \emph{IEEE International Conference on Robotics and
  Automation, 2004. Proceedings. ICRA '04. 2004}, vol.~4, 2004, pp. 3512--3518
  Vol.4.

\bibitem{Li2007}
Y.~Li, J.~L. Fu, and N.~S. Pollard, ``Data-driven grasp synthesis using shape
  matching and task-based pruning,'' \emph{IEEE Transactions on Visualization
  and Computer Graphics}, vol.~13, no.~4, pp. 732--747, 2007.

\bibitem{Tian2019}
H.~Tian, C.~Wang, D.~Manocha, and X.~Zhang, ``Realtime hand-object interaction
  using learned grasp space for virtual environments,'' \emph{IEEE Transactions
  on Visualization and Computer Graphics}, vol.~25, no.~8, pp. 2623--2635,
  2019.

\bibitem{Kleeberger2020}
\BIBentryALTinterwordspacing
K.~Kleeberger, R.~Bormann, W.~Kraus, and M.~F. Huber, ``A survey on
  learning-based robotic grasping,'' \emph{Current Robotics Reports}, vol.~1,
  no.~4, pp. 239--249, Dec 2020. [Online]. Available:
  \url{https://doi.org/10.1007/s43154-020-00021-6}
\BIBentrySTDinterwordspacing

\bibitem{Souza2021}
\BIBentryALTinterwordspacing
J.~P.~C. {de Souza}, L.~F. Rocha, P.~M. Oliveira, A.~P. Moreira, and
  J.~Boaventura-Cunha, ``Robotic grasping: from wrench space heuristics to deep
  learning policies,'' \emph{Robotics and Computer-Integrated Manufacturing},
  vol.~71, p. 102176, 2021. [Online]. Available:
  \url{https://www.sciencedirect.com/science/article/pii/S0736584521000594}
\BIBentrySTDinterwordspacing

\bibitem{Sahbani2012}
\BIBentryALTinterwordspacing
A.~Sahbani, S.~El-Khoury, and P.~Bidaud, ``An overview of 3d object grasp
  synthesis algorithms,'' \emph{Robotics and Autonomous Systems}, vol.~60,
  no.~3, pp. 326--336, 2012, autonomous Grasping. [Online]. Available:
  \url{https://www.sciencedirect.com/science/article/pii/S0921889011001485}
\BIBentrySTDinterwordspacing

\bibitem{Nguyen1986}
V.-D. Nguyen, ``Constructing force-closure grasps,'' in \emph{Proceedings. 1986
  IEEE International Conference on Robotics and Automation}, vol.~3, 1986, pp.
  1368--1373.

\bibitem{Bicchi2000}
A.~Bicchi and V.~Kumar, ``Robotic grasping and contact: a review,'' in
  \emph{Proceedings 2000 ICRA. Millennium Conference. IEEE International
  Conference on Robotics and Automation. Symposia Proceedings (Cat.
  No.00CH37065)}, vol.~1, 2000, pp. 348--353 vol.1.

\bibitem{Wan2012}
W.~Wan, R.~Fukui, M.~Shimosaka, T.~Sato, and Y.~Kuniyoshi, ``Grasping by
  caging: A promising tool to deal with uncertainty,'' in \emph{2012 IEEE
  International Conference on Robotics and Automation}, 2012, pp. 5142--5149.

\bibitem{Kim2019}
\BIBentryALTinterwordspacing
D.~Kim, Y.~Maeda, and S.~Komiyama, ``Caging-based grasping of deformable
  objects for geometry-based robotic manipulation,'' \emph{ROBOMECH Journal},
  vol.~6, no.~1, p.~3, Mar 2019. [Online]. Available:
  \url{https://doi.org/10.1186/s40648-019-0131-4}
\BIBentrySTDinterwordspacing

\bibitem{Su2017}
J.~Su, H.~Qiao, C.~Liu, Y.~Song, and A.~Yang, ``Grasping objects: The
  relationship between the cage and the form-closure grasp,'' \emph{IEEE
  Robotics \& Automation Magazine}, vol.~24, no.~3, pp. 84--96, 2017.

\bibitem{Laguna2013}
\BIBentryALTinterwordspacing
M.~Laguna and R.~Mart{\'i}, \emph{Heuristics}.\hskip 1em plus 0.5em minus
  0.4em\relax Boston, MA: Springer US, 2013, pp. 695--703. [Online]. Available:
  \url{https://doi.org/10.1007/978-1-4419-1153-7_1184}
\BIBentrySTDinterwordspacing

\bibitem{Bohg2014}
J.~Bohg, A.~Morales, T.~Asfour, and D.~Kragic, ``Data-driven grasp
  synthesis—a survey,'' \emph{IEEE Transactions on Robotics}, vol.~30, no.~2,
  pp. 289--309, 2014.

\bibitem{Aleotti2012}
J.~Aleotti and S.~Caselli, ``Grasp programming by demonstration in virtual
  reality with automatic environment reconstruction,'' \emph{Virtual Real.},
  vol.~16, no.~2, p. 87–104, jun 2012.

\bibitem{Saxena2008}
\BIBentryALTinterwordspacing
A.~Saxena, J.~Driemeyer, and A.~Y. Ng, ``Robotic grasping of novel objects
  using vision,'' \emph{The International Journal of Robotics Research},
  vol.~27, no.~2, pp. 157--173, 2008. [Online]. Available:
  \url{https://doi.org/10.1177/0278364907087172}
\BIBentrySTDinterwordspacing

\bibitem{Zaidi2017}
\BIBentryALTinterwordspacing
L.~Zaidi, J.~A. Corrales, B.~C. Bouzgarrou, Y.~Mezouar, and L.~Sabourin,
  ``Model-based strategy for grasping 3d deformable objects using a
  multi-fingered robotic hand,'' \emph{Robotics and Autonomous Systems},
  vol.~95, pp. 196--206, 2017. [Online]. Available:
  \url{https://www.sciencedirect.com/science/article/pii/S0921889016308089}
\BIBentrySTDinterwordspacing

\bibitem{Deformation-xi}
{Wikipedia contributors}, ``Deformation (engineering),''
  \url{https://en.wikipedia.org/w/index.php?title=Deformation_(engineering)\&
  oldid=1189053730}, Dec. 2023, accessed: NA-NA-NA.

\bibitem{Yu2023}
\BIBentryALTinterwordspacing
X.~Yu, R.~Huang, C.~Zhao, L.~Zhou, and L.~Ou, ``Defgrasp: A robot grasping
  detection method for deformable objects without force sensor,'' \emph{Neural
  Processing Letters}, vol.~55, no.~8, p. 11739–11756, 2023. [Online].
  Available: \url{https://doi.org/10.1007/s11063023113988}
\BIBentrySTDinterwordspacing

\bibitem{Xu2020}
J.~Xu, M.~Danielczuk, J.~Ichnowski, J.~Mahler, E.~Steinbach, and K.~Goldberg,
  ``Minimal work: A grasp quality metric for deformable hollow objects,'' in
  \emph{2020 IEEE International Conference on Robotics and Automation (ICRA)},
  2020, pp. 1546--1552.

\bibitem{Adachi2015}
R.~Adachi, Y.~Fujihira, and T.~Watanabe, ``Identification of danger state for
  grasping delicate tofu with fingertips containing viscoelastic fluid,'' in
  \emph{2015 IEEE/RSJ International Conference on Intelligent Robots and
  Systems (IROS)}, 2015, pp. 497--503.

\bibitem{Nishimura2016}
T.~Nishimura, Y.~Fujihira, R.~Adachi, and T.~Watanabe, ``New condition for tofu
  stable grasping with fluid fingertips,'' in \emph{2016 IEEE International
  Conference on Automation Science and Engineering (CASE)}, 2016, pp. 335--341.

\bibitem{SU2019}
\BIBentryALTinterwordspacing
J.~Su, B.~Chen, H.~Qiao, and Z.~yong Liu, ``Caging a novel object using
  multi-task learning method,'' \emph{Neurocomputing}, vol. 351, pp. 146--155,
  2019. [Online]. Available:
  \url{https://www.sciencedirect.com/science/article/pii/S0925231219304035}
\BIBentrySTDinterwordspacing

\bibitem{Grimm20221}
\BIBentryALTinterwordspacing
P.~Grimm, W.~Broll, R.~Herold, J.~Hummel, and R.~Kruse, \emph{VR/AR Input
  Devices and Tracking}.\hskip 1em plus 0.5em minus 0.4em\relax Cham: Springer
  International Publishing, 2022, pp. 107--148. [Online]. Available:
  \url{https://doi.org/10.1007/978-3-030-79062-2_4}
\BIBentrySTDinterwordspacing

\bibitem{Sharp2015}
\BIBentryALTinterwordspacing
T.~Sharp, C.~Keskin, D.~Robertson, J.~Taylor, J.~Shotton, D.~Kim, C.~Rhemann,
  I.~Leichter, A.~Vinnikov, Y.~Wei, D.~Freedman, P.~Kohli, E.~Krupka,
  A.~Fitzgibbon, and S.~Izadi, ``Accurate, robust, and flexible real-time hand
  tracking,'' in \emph{Proceedings of the 33rd Annual ACM Conference on Human
  Factors in Computing Systems}, ser. CHI '15.\hskip 1em plus 0.5em minus
  0.4em\relax New York, NY, USA: Association for Computing Machinery, 2015, p.
  3633–3642. [Online]. Available:
  \url{https://doi.org/10.1145/2702123.2702179}
\BIBentrySTDinterwordspacing

\bibitem{Openaai}
``\BIBforeignlanguage{en}{Openarcloud},'' \url{https://www.openarcloud.org/},
  accessed: 2024-6-20.

\bibitem{Clay2019-lr}
V.~Clay, P.~K{\"o}nig, and S.~K{\"o}nig, ``\BIBforeignlanguage{en}{Eye tracking
  in virtual reality},'' \emph{\BIBforeignlanguage{en}{J. Eye Mov. Res.}},
  vol.~12, no.~1, Apr. 2019.

\end{thebibliography}
\end{document}